\documentclass[aps,12pt,longbibliography]{revtex4-1}

\usepackage[lofdepth,lotdepth]{subfig}
\usepackage{graphicx}
\usepackage{algorithm}
\usepackage{algpseudocode}
\usepackage{amssymb}
\usepackage{gensymb}

\usepackage[latin1]{inputenc}
\usepackage{tikz}
\usetikzlibrary{shapes,arrows,positioning,calc}

\usepackage{ulem}

\newcommand{\R}[1]{{\color{black}#1}}

\newcommand{\B}[1]{{\color{blue}#1}}

\newcommand{\vJ}{{\bf J}}

\newcommand{\vB}{{\bf B}}

\newcommand{\vH}{{\bf H}}
\newcommand{\vT}{{\bf T}}

\tikzstyle{startstop} = [rectangle, rounded corners, minimum width=3cm, minimum height=1cm,text centered, draw=black, fill=red!30]
\tikzstyle{io} = [trapezium, trapezium left angle=70, trapezium right angle=110, minimum width=3cm, minimum height=1cm, text centered, draw=black, fill=blue!30]
\tikzstyle{process} = [rectangle, minimum width=3cm, minimum height=1cm, align=left, draw=black, fill=orange!30]
\tikzstyle{empty} = [rectangle, align=left]
\tikzstyle{decision} = [diamond, minimum width=3cm, minimum height=1cm, align=center, draw=black, fill=green!30]
\tikzstyle{arrow} = [thick,->,>=stealth]

\begin{document}

\title{Cross-field demagnetization of stacks of tapes: 3D modelling and measurements\footnote{\B{This article has been published in Supercond. Sci. Technol. with doi:10.1088/1361-6668/ab5aca}}} %
\markboth{Cross-field demagnetization of the stack of tapes ...}{}

\author{M. Kapolka$^1$, E. Pardo$^{1}$\footnote{Corresponding author: enric.pardo@savba.sk}, F. Grilli$^2$, A. Baskys$^{3,4}$, V. Climente-Alarcon$^3$\R{, A. Dadhich$^{1}$}, B. A. Glowacki$^{3,5}$}%

\address{$^1$Institute of Electrical Engineering, Slovak Academy of Sciences, Dubravska 9, 84104 Bratislava, Slovakia}
\address{$^2$Institute for Technical Physics, Karlsruhe Institute of Technology, 76131 Karlsruhe, Germany}
\address{$^3$Applied Superconductivity and Cryoscience Group, Department of Materials Science and Metallurgy, University of Cambridge, 27 Charles Babbage Road, Cambridge, CB3 0FS, United Kingdom}
\address{$^4$European Organization for Nuclear Research (CERN), 1211 Geneva 23, Switzerland}
\address{$^5$Institute of Power Engineering, ul. Mory 8, 01-330 Warsaw, Poland}

\date{\today}

\begin{abstract}

Stacks of superconducting tapes can trap much higher magnetic fields than conventional magnets. This makes them very promising for motors and generators. However, ripple magnetic fields in these machines present a cross-field component that demagnetizes the stacks. At present, there is no quantitative agreement between measurements and modeling \R{of cross-field demagnetization}, mainly due to the need of a 3D model that takes the \R{end effects and} real micron-thick superconducting layer into account. This article presents 3D modeling and measurements of cross-field demagnetization in stacks of up to 5 tapes and initial magnetization modeling of stacks of up to 15 tapes. 3D modeling of the cross-field demagnetization \R{explicitly} shows that the critical current density, $J_c$, in the \R{direction perpendicular to the tape surface} does not play a role in cross-field demagnetization. When taking the measured anisotropic magnetic field dependence of $J_c$ into account, \R{3D} calculations agree with measurements with less than 4 \% deviation\R{, while the error of 2D modeling is much higher}. Then, our 3D numerical methods can realistically predict cross-field demagnetization. Due to the force-free configuration of part of the current density, $J$, in the stack, better agreement with experiments will \R{probably} require \R{measuring} the $J_c$ anisotropy for the whole solid angle range, including $J$ parallel to the magnetic field.

\end{abstract}

\maketitle


\section{Introduction}
\label{s.intro}

Stacks of superconducting \R{REBCO} tapes after magnetization behave like permanent magnets \R{but with superior trapped field, being the world record of 17.7 T \cite{Patel18SST}} compared to around 1.3 T \R{maximum} remnant magnetic field of conventional permanent magnets. Although superconducting bulks can also trap high magnetic fields (17.6 T \cite{Durrell14SST}), stacks present additional advantages. Their Hastelloy substrate enhances mechanical properties. \R{Also, the} stack lenght is virtually unlimited with very uniform $J_c$, and the stack width could be as wide as 46 mm \cite{Rupich16IES}. \R{Larger} continuous superconducting \R{shapes result} in larger trapped flux for the same maximum trapped field. \R{Following the critical-state model (CSM), } a \R{saturated} bulk mosaic made of hexagonal or square tiles traps an average flux density of around $1/3$ of its maximum, while the average flux density on an stack is around $1/2$ of its maximum \R{\cite{chen89JAP}}. Then, a long stack traps 50 $\%$ more flux than an array of bulks of the same width as the stack for each bulk. \R{In addition,} the stack enables to interlay sheets of other materials to enhance physical properties\R{:} metal layers enhance thermal properties\R{; and} soft ferromagnetic layers enhance \R{the} trapped field and reduce cross-field demagnetization, at least for stacks as stand-alone objects \R{and below the saturation for the magnetic material} \cite{Baghdadi18SR}.     

\R{The high trapped flux in superconducting stacks can be exploited to enhance the magnetic flux density at the gap of motors and generators, when placing these materials in the rotor \cite{patel18IEE,climentealarcon19IES,sotelo18IIE}. Other alternatives to achieve the same goal by means of superconductors is to use bulks \cite{zhouD12SST,ainslie15SST,Yanamoto17IES,shaanika19IES} or pole coils in the rotor \cite{Gamble11IES,bergen19SST}. Higher gap flux densities enable weigh and size reduction for the same power and torque ratings. This feature can be further enhanced by adding a superconducting stator \cite{oswald12PhP,Pardo19IES,eucas19motor}, resulting in a full superconducting motor. Thanks to this, superconducting machines present a high potential} \cite{Haran17SST}, especially for electric aircrafts \R{\cite{masson13IES,Berg17IES,Asumed,patel18IEE,Pardo19IES,eucas19motor,corduan19prp}}, high power generators \cite{marino16SST,bergen19SST,corduan19prp} and sea transport \cite{Gamble11IES,Yanamoto17IES,shaanika19IES}. 

\R{For the stack of tapes technology, the ripple transverse fields that the stacks experience in the rotor cause demagnetization of the trapped field, being a major issue. The acceptable level of demagnetization depends on the application. For aircraft propulsion, the rated power should be kept during commercial flights, of a few hours. One option could be that at take-off the stacks are over-magnetized to trapped fields above the requirements, in order to guarantee the rated level throughout the flight. Then, we can estimate that demagnetizations above 30 \% in a few hours are impractical.}

There \R{has been} a big effort to fully understand the cross-field demagnetization, in order to reduce demagnetization \R{effects} and extend the time of the trapped field inside the stack, as follows.  

Recent measurements of stacks showed the \R{main} behaviour under cross and rotating magnetic fields \R{\cite{Baghdadi14APL, Campbell17SST, Baghdadi18SR, Baskys18SST}}, \R{the latter reporting measurements for up to 10~000 cycles.} However, theoretical study by cross-sectional approximation (\R{infinitely long} 2D approach) showed qualitative agreement only \R{\cite{Campbell17SST, Baghdadi18SR}}. \R{In addition,} \cite{Campbell17SST} \R{compares an} $A$-formulation with Brandt and Mikitik theory \cite{Brandt02PRL}\R{, showing good agreement}. \R{One reason of quantitative disagreement with experiments is the unrealistically high thickness in the models, being 10-20 $\mu$m compared to 1-2 $\mu$m in experimental samples. Liang et al. take the real thickness into account in their 2D modeling, showing again good qualitative agreement but still quantitative discrepancies \cite{Liang17SST}. As stated in \cite{Liang17SST}, the remaining discrepancy is due to the end effects of the relatively short experimental samples (usually made of square tapes), requiring 3D modeling.} 3D models \R{have been only published} for cylindrical bulks by Fagnard et al \cite{Fagnard16SST} or cubic bulks by Kapolka et al \cite{Kapolka18IES}. However, in the overall the full 3D model of the stack of tapes is missing, where good qualitative agreement with experiments is expected. \R{In addition, 3D modeling can enable to study also the current density component parallel to the ripple field; in contrast to 2D modeling, which can describe the perpendicular component only}.

The main reason of missing 3D models is due to the low superconducting thickness of around 1 $\mu$m \R{and the need to mesh several elements across the thickness}, \R{resulting in a} high aspect ratio of the elements. Since the variation of current density across the thickness is essential for cross-field demagnetization, methods assuming the thin-film approach cannot be applied. The inaccuracy of the elongated elements leads to numerical issues such as high number of elements, instability and the non-convergence of the modeling tools. \R{These are the reasons why} models often do not take the real thickness of the superconducting layer into account. 

Our goal is to model Stacks with the real thickness of the superconducting layer 1.5 $\mu$m and compare the results to the measurements, being the first 3D model of the cross-field demagnetization of stacks of tapes. \R{Thus, following the methodology definitions in \cite{stenvall19IES}, we ``attack a well-known problem at the frontiers of knowledge".} For the \R{studied configuration}, our method (MEMEP 3D) is more efficient than (FEM) in $\vH$ formulation, because, due to the thin film shape, FEM uses many elements in the air around the sample and even between thin films \cite{grilli13Cry}. \R{In addition, certain software packages like COMSOL present issues for 3D elements with the required high aspect ratio. The stack cross-feild configuration} also seems not suitable for Fast Fourier Transformation (FFT). The bulk FFT approach \cite{Prigozhin18SSTa} requires a cumbersome number of elements due to the need of uniform mesh and the low film thickness. The stack approach of FFT assumes thin films for the tapes \cite{Prigozhin18SST}, which cannot describe cross-field demagnetization. \R{Then, we model the experimental geometry by MEMEP 3D. Another advantage of MEMEP 3D is that it is able to take macroscopic force-free effects into account \cite{Kapolka19SST}, backed by the theory of Badia and Lopez \cite{badia15SST}.}

In this article we focus on the cross-field demagnetization of the stacks of tapes up to 5 tapes \R{with MEMEP 3D}, the validation of our (MEMEP 3D) method by comparison of two-tapes demagnetization with FEM, the trapped field in the stack up to 15 tapes and qualitative behavior of \R{bulks and stacks} with similar parameters.


\section{Methodology of measurements}
\label{s.methodology}

The study is focused on the cross-field demagnetization of a stack of tapes. The sample is prepared from 12 mm wide SuperOx tapes with stated minimum $I_c$ of 430 A at 77 K. The thickness of the tape is around 65 $\mu$m with 1.5 $\mu$m thin superconducting (SC) layer. The tape is with $\sim$ 2 $\mu$m silver stabilization on each side and around 60 $\mu$m Hastelloy. The stack of tapes is formed by 5 SuperOx tapes with 3 Kapton layers between each superconducting layer. The superconducting tape together with 3 Kapton insulators is 220 $\mu$m thick. The sensitive part of the Hall probe sensor is 1.5 mm above the top SC layer. \R{The sensitivity of the probe is at least 10 mV/T}.      

Cross-field demagnetization consists on the following three main steps: magnetization by field \R{cooling} (FC) method, relaxation time and cross-field demagnetization. The detailed process is the following:
\begin{itemize}
	\item The sample is placed into the electromagnet at room temperature.
	\item The electromagnet is ramped up to 1 T.  
	\item The sample is cooled down in liquid nitrogen bath at 77 K.
	\item The electromagnet is ramped down with ramp rate 10 mT/s. 
	\item The sample is moved into the air-core solenoid. 
	\item The sample is left for 300 s relaxation.
	\item \R{The Arepoc Hall sensor LHP-MPc is placed 1.5 mm} above the sample measured the trapped magnetic field $B_t$ at the center.
	\item The solenoid magnet applied a sinusoidal transverse-(or cross) magnetic field of several amplitudes (50, 100, 150 mT) and frequencies (1, 10 Hz). The Hall probe measured the trapped field during the demagnetization.
\end{itemize}

\begin{figure}[tbp]
\centering
{\includegraphics[trim=0 0 0 0,clip,width=6 cm]{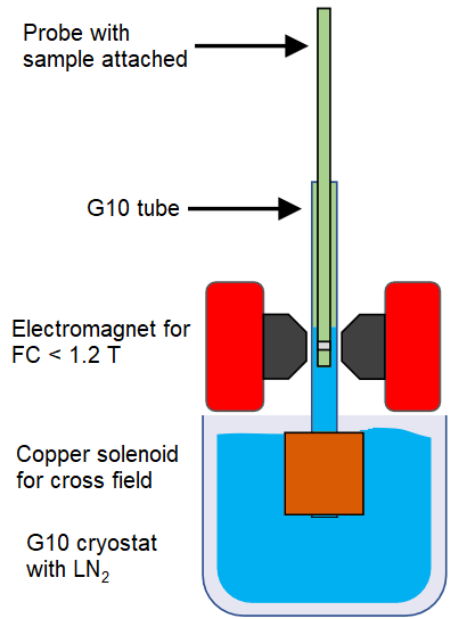}}
\caption{\R{The measurement set-up magnetizes the superconducting stack by field cooling using an electromagnet and later applies alternating cross-field demagnetization by means of a copper coil.}}
\label{setup.fig}
\end{figure}

The measurement set-up contains a G10-cryostat for sample holder, an iron-core Walker Scientific HV-4H electromagnet and the separated air core solenoid \R{(figure \ref{setup.fig})}. 
 
The control system \cite{Baskys18SST} contains a signal generator Agilent 33220A amplified by two power supplies KEPCO BOP 2020 connected in parallel. The current in the circuit is measured by a LEM Ultrastab IT 405-S current transducer. The magnetic field is measured by an Arepoc Hall sensor LHP-MPc \cite{Arepoc}. The circuit is monitored by custom made lab-View program. 


\section{Modelling method}

In this article, we use two different numerical methods. Most of the calculations are made with the Minimum Electro-Magnetic Entropy Production method in 3D (MEMEP 3D), although we first benchmarked this method with Finite Element Method (FEM) calculation in the $\vH$ formulation for simple cases in order to cross-check the numerical methods. Although, in the previous work we made a validation of the three methods by magnetization of the bulk and stacks with the tilted fields \cite{Kapolka18IESa}.

\subsection{Modelling conditions}

The SuperOx tape contains Hastelloy, superconducting layer (SC) and the silver thin layer as it is explained in section \ref{s.methodology}. However, the model takes only the superconducting layer into account. Then, the Hastelloy, silver \R{and cooling medium} are treated as void (or ``air") in the model, \R{since these materials and non-magnetic and (for the metals) their eddy currents are negligible in the studied frequency range (up to 500 Hz)}. \R{From now on, we refer to ``tape" as the superconducting layer only.} The gap is around 200 $\mu$m, being slightly different for each studied configuration. The thickness of the SC layer depends on the goal of the study. The general qualitative study of the cross-field demagnetization uses the thickness of 10 $\mu$m and the more precise calculation for comparison to experiments uses the real thickness of 1.5 $\mu$m. \R{Unless stated otherwise, the frequency of the ripples is taken as 500 Hz. We chose this characteristic frequency because the rotor in electric machines for aviation presents ripple fields of fundamental frequencies of few hundreds of hertz or higher, since their rotating speed is targeted to few thousands of rpm \cite{patel18IEE,corduan19prp}.}

\subsection{MEMEP 3D model}

We perform most of the calculations here by MEMEP 3D \cite{Pardo17JCP} based on a variational principle\R{, being the real-thickness calculations and comparison to experiments done by this method only}. The mathematical formulation uses the $\vT$ vector defined as an effective magnetization. The effective magnetization is non-zero only inside the modelling sample, and hence the method does not solve the \R{surrounding} air domain. \R{However, the air separation between superconducting layers in the stack is modelled as a conducting material of high resistivity.} The incorporated isotropic power-law enables to take $n$ values up to $n$=1000 into account. We use $n$=200 as an approximation to the Critical State Model and $n$=30 as a realistic value for the measurements. The modelling software \cite{Pardo17SST} was developed in C++ and it is enhanced by parallel computation on a computer cluster \R{\cite{Kapolka19SST,kapolka18PhD,Cluster}}. The method uses hexahedric elements with high aspect ratio, up to 5000. Therefore, MEMEP can use the same modelling geometry as the measured samples. The model assumed either isotropic constant $J_c$, $J_c(B)$ or $J_c(B,\theta)$ from measurements, depending on the configuration, being $B$ the magnitude of the magnetic field (we use the term ``magnetic field" for both $\vB$ and $\vH$, since for our case $\vB=\mu_0\vH$) and $\theta$ is the angle between $\vB$ and the normal of the tape surface [figure \ref{JcB_data.fig} (b)]. \R{Some calculations are done in the 2D version of MEMEP \cite{pardo15SST}, to assess the finite-sample effects. All modeling results are 3D, unless stated otherwise.}

\begin{figure}[tbp]
\centering
{\includegraphics[trim=0 0 0 0,clip,width=12 cm]{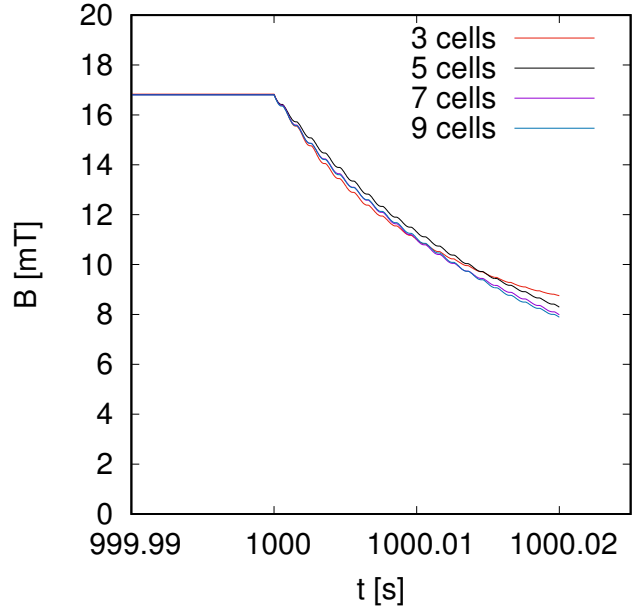}}
\caption{\R{The modeling results of cross-field demagnetization are mesh independent for the used mesh of $15\times 15\times 9$ elements in each tape. Results above for a single tape with several number of cells in the thickness, ripple field of 240 mT amplitude and 500 Hz frequency, 10 cycles of ripple field, and field cooling process shown in figure \ref{wave_form.fig} followed by 900 s of relaxation. Cross-field demagnetization starts at time 1000~s.}}
\label{meshdep.fig}
\end{figure}

\begin{figure}[tbp]
\centering
\subfloat[][]
{\includegraphics[trim=10 0 10 0,clip,width=8 cm]{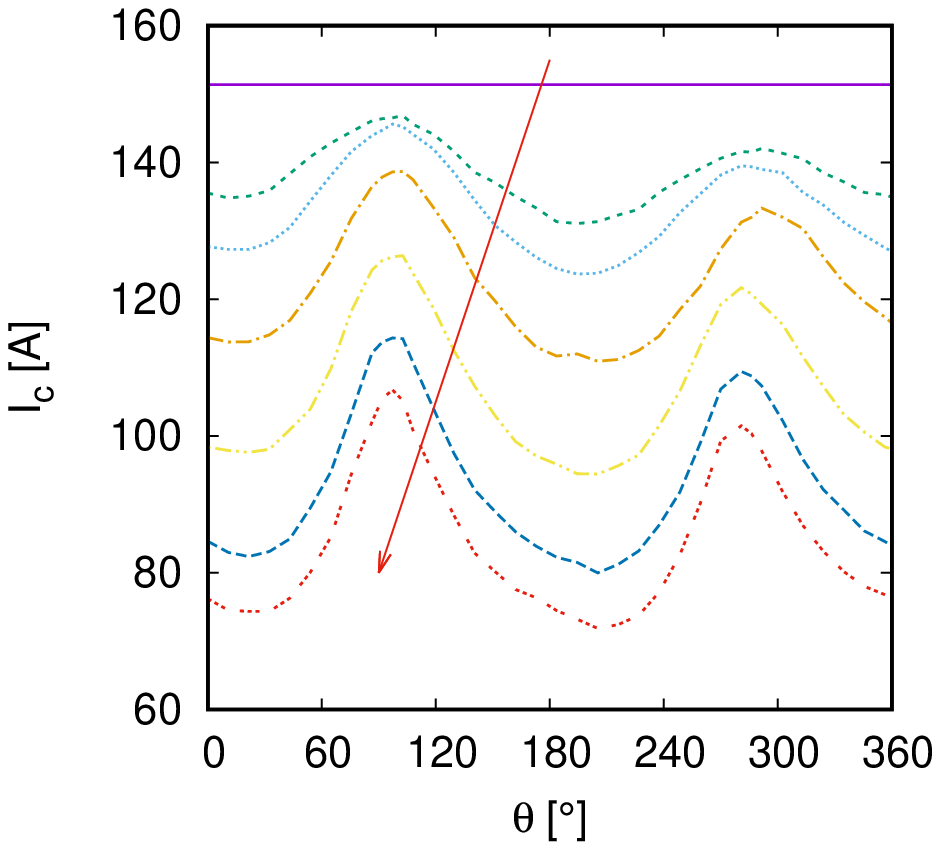}}
\subfloat[][]
{\includegraphics[trim=0 0 0 0,clip,width=8 cm]{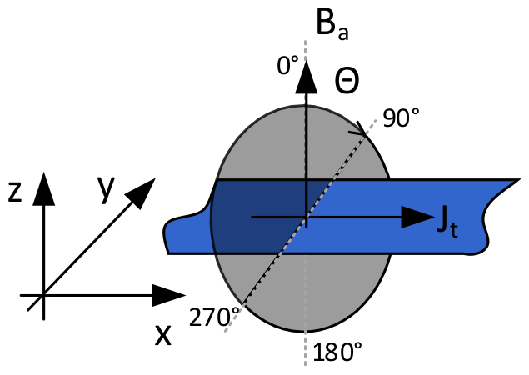}}
\caption{(a) The $I_c(B,\theta)$ measured data on a 4 mm wide SuperOx tape \R{for} the applied \R{fields} $B_{a}$=0, 36, 49, 72, 100, 144, 180 mT \R{in the arrow direction, using the set-up in Bratislava \cite{seiler19SST}}. (b) Sketch of the $I_c(B, \theta)$ measurements with the definition of the $\theta$ angle to the tape surface.}
\label{JcB_data.fig}
\end{figure}

\R{Unless stated otherwise, the number of elements in the superconductor are $15\times 15\times 9$ per tape, being the last the number of elements in the thickness. We have checked that with this mesh, the calculated cross-field demagnetization is mesh-independent. As seen in figure \ref{meshdep.fig}, the difference in trapped field in one tape for both at the end of relaxation and after 10 cycles is the same for 7 and 9 cells across the thickness. We also checked that the results are insensitive to the number of cells in the tape width. and  This is the mesh used for the comparison with experiments. For all calculations, we use a tolerance for $J$ of 0.01 \% of $J_c$ at zero magnetic field, although we checked that the results do not differ for a tolerance of 0.001 \%.}

\R{The variational formulation of MEMEP has been extended to take magnetic materials into account, as shown in the 2D examples of \cite{pardo16HTSmod}. However, magnetic materials are not yet implemented in the 3D version of the software. This is not an issue in this article because the studied tape substrate is non-magnetic.} 

\subsection{FEM 3D model}

{Here, we use FEM in order to benchmark the MEMEP 3D method.} The FEM model is based on the $H$-formulation of Maxwell's equations implemented in the finite-element program Comsol Multiphysics \cite{brambilla07SST}. Due to the necessity of simulating the air between and around the superconducting tapes typical of the FEM approach, care had to be taken in building the domains and the mesh. In order to avoid an excessive number of degrees of freedom, an approach based on sweeping a 2D geometry and mesh was followed (see Fig. 1 of \cite{grilli13Cry} for an example). The external magnetic field was applied on the boundary of the air domains by means of Dirichlet boundary conditions. A magnetic field of $B_z$=300 mT was assigned to all simulated domains as initial condition. \R{For FEM, we use $15\times 15\times 5$ elements in each superconducting tape and a tolerance of 0.5 \%.}


\section{Results and discussion}

There is a big effort to fully understand the cross-field demagnetization process. However, almost all studies found only qualitative agreement with measurements. We focused on 3D modelling with all finite size effects, and hence the results can be compared with measurements on short samples. For the comparison to experiments, the model assumes the real dimensions of the measured sample and measured $J_c(B,\theta)$ dependence. Before comparing to experiments, we analyse the influence of several parameters like the superconducting layer thickness and gap between tapes. For this analysis, we study first the magnetization process and later the cross-field demagnetization.


\subsection{In-plane magnetization of single tape with thickness variation }
\label{s.in_plane}

\begin{figure}[tbp]
\centering
{\includegraphics[trim=0 0 0 0,clip,height=4.5 cm]{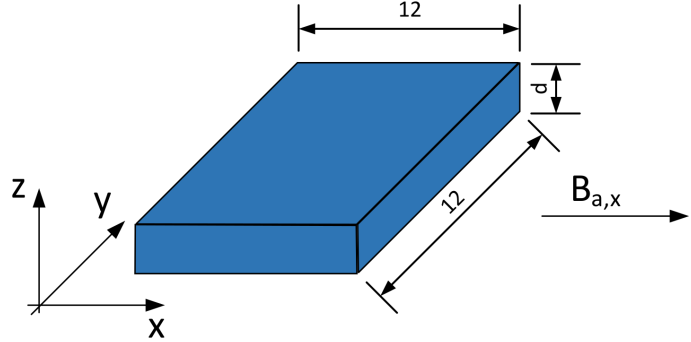}}
\caption{Single superconducting tape for the thickness dependence study ($d$=1,10,100,1000 $\mu$m). Dimensions are in mm.}
\label{sketch1.fig}
\end{figure}

The first study is about in-plane magnetization due to a parallel applied magnetic field. The sketch of the modelling case and the dimensions are in the figure \ref{sketch1.fig}. The magnetization loop is calculated only for the $M_x$ component, because the applied magnetic field (of amplitude 50 mT and 50 Hz frequency) is along the $x$ axis. We used a thicknesses of the superconducting layer $d$ in the range from 1 to 1000 $\mu$m for this case. We assume constant $J_cd$, being $J_c$=2.72$\times 10^{10}$ A/m$^2$ for $d$=1 $\mu$m and $n$-power law exponent of 30.

\begin{figure}[tbp]
\centering
 \subfloat[][]
{\includegraphics[trim=69 0 70 0,clip,height=7.0 cm]{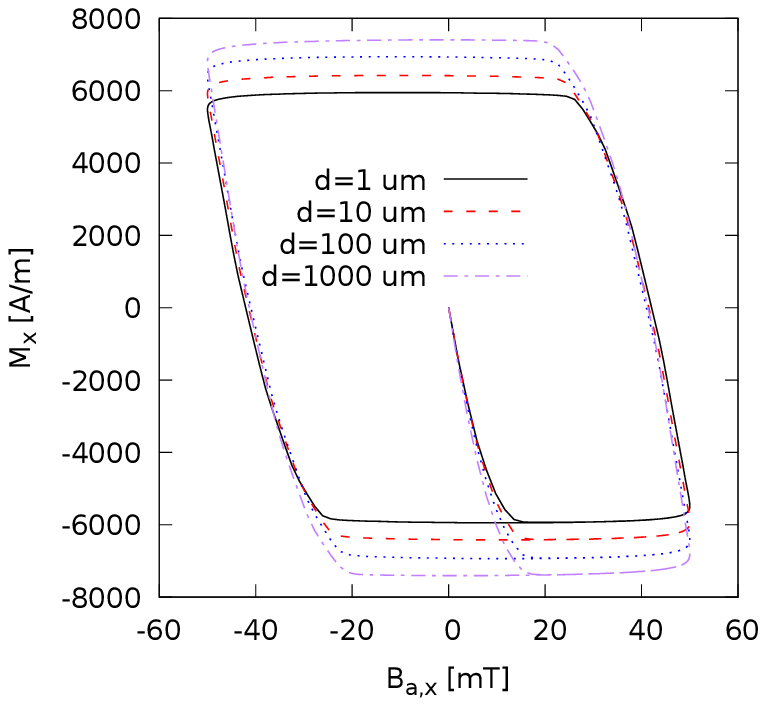}}
 \subfloat[][]
{\includegraphics[trim=0 0 0 0,clip,height=7.0 cm]{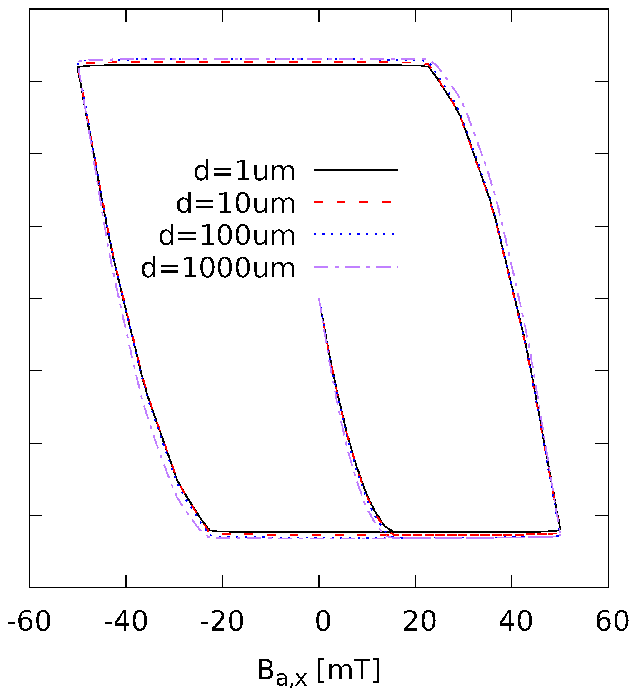}}  
\caption{Magnetization loops $M_x$ of the single tape increases with thickness $d$=(1,10,100,1000 $\mu$m), while keeping the sheet critical current density $J_cd$ constant. The model assumes a realistic $n$ power law exponent (a) n=30 and a situation close to the Critical State Model (b) n=200.}
\label{mag_loop.fig}
\end{figure}

The magnetization increases with tape thickness (Fig. \ref{mag_loop.fig} (a)). This is not the case for the Critical State Model (CSM), which we approximate as a power law with exponent $n=200$ (Fig. \ref{mag_loop.fig} (a)). All magnetization loops for different thickness are within 2$\%$, being not identical due to the power-law exponent (Fig. \ref{mag_loop.fig} (b)). However, commercial superconducting tapes have an $n$-value around 30 in self-field. The thickness dependence is due to higher electric fields from the applied magnetic field in thicker tapes. The relatively low $n$-value of 30 allows $J>J_c$ for high \R{local} electric fields, and hence the magnetization increases with thickness roughly as 14$\%$ for each increase in thickness by a factor 10. \R{This effect does not appear for the CSM, since $|\vJ|$ is limited to $J_c$, irrespective to the value of the non-zero electric field.} Therefore, we already see that the sample thickness plays a role for the response to the ripple field. The effects of the thickness is much more important for the cross-field demagnetization (section \ref{s.thickness}), which is also significant for the CSM \cite{Brandt02PRL, Brandt04PhC}. In the calculations in the following sections, we assume the real thickness of 1.5 $\mu$m for comparison to experiments and 10 $\mu$m for the purely numerical systematic analysis.     

\subsection{Cross-field demagnetization of one tape with thickness variation}
\label{s.thickness}

A more detailed study about thickness influence is in the cross-field demagnetization. The model assumes a single tape with thickness from 1 to 100 $\mu$m. \R{Here, we make a similar analysis as for previous 2D modeling starting from 10 $\mu$m upwards for a single cycle \cite{Baghdadi18SR} but now for the 3D configuration, starting from 1 $\mu$m, and up to 10 cycles, being more relevant for experiments.} The critical current density is inversely proportional to the thickness. The $J_c$ of 1 $\mu$m tape is $J_c$=2.72$\times 10^{10}$ A/m$^2$ and the $n$ value is 30. The tape is magnetized with the perpendicular applied field to the tape surface by the  field \R{cooling} method. The field is ramped down with rate 30 mT/s over 100 s with following relaxation of 900 s. Afterwards, a sinusoidal transverse field of 500 Hz is applied along the $x$ axis. 

\begin{figure}[tbp]
\centering
{\includegraphics[trim=0 0 0 0,clip,height=7.0 cm]{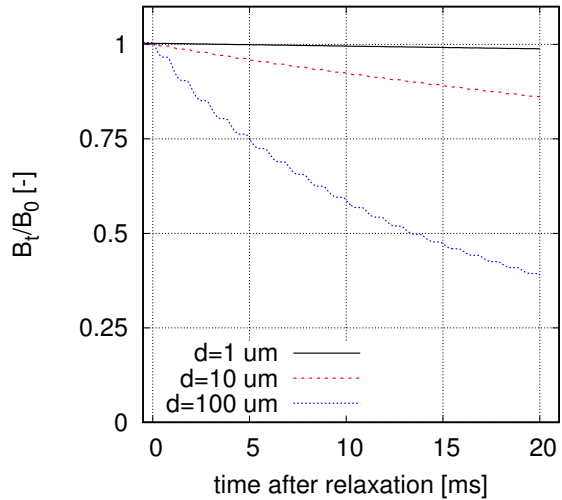}}
\caption{\R{For constant $J_cd$, being $d$ the tape thickness, the trapped field, $B_t$, decreases with the thickness, and hence the model needs to take the real tape thickness of the order of 1 $\mu$m into account. Calculations in this graph are for 1 mm above a single tape, power-law exponent $n$=30 and normalized to the trapped field at the end of relaxation, $B_0$.}} 
\label{thickness.fig}
\end{figure}

The demagnetization rate significantly increases with the thickness (figure \ref{thickness.fig}), even though $J_c$ is proportional to the thickness. The clear thickness dependence showed the importance of the sample thickness, as it is explained in section \ref{s.in_plane}. Therefore, the thickness in the model is very important for the cross-field demagnetization, where ripples are in the in-plane direction. The model cannot assume thicker films and lower proportionally the critical current density, as already predicted by Campbell et al \cite{Campbell17SST}. Since the superconducting layer in most REBCO tapes is of the order of 1 $\mu$m, 3D modelling is very challenging due to the high aspect ratio. Most previous works, which are in 2D, assumed unrealistically thicker samples due to numerical issues.


\subsection{Trapped magnetic field in the stack of tapes}

The next study is about the influence of number of tapes and gap between superconducting layers in the stack on the initial trapped field. We used the same geometrical parameters as in the previous section \ref{s.in_plane}. The superconducting layer is 10 $\mu m$ thick with $J_c$= 2.72$\times 10^9$ A/m$^2$ and $n$=30. As shown above, it is not possible to take a larger superconductor thickness and a proportionally lower $J_c$ for cross-field demagnetization and transverse applied field. However, this simplification could be made for applied fields perpendicular to the tapes \cite{Pardo12SSTa,acreview}. The cause is that the electric field due to the applied magnetic field is roughly uniform in the tape thickness and the effect of the self-magnetic field can be averaged over the tape thickness. 

The stack is magnetized by field \R{cooling} (FC) method along the $z$ axis. The initial applied magnetic field is 1 T with ramp down rate of 10 mT/s, because of the perpendicular penetration field $B_p$ of 1 tape is 27.2 mT. Afterwards, we leave a relaxation time of 900 s, which is long enough to reach stable state of the trapped field. The trapped field decreases logarithmically during relaxation, and hence after a short time the reduction is almost negligible. The trapped field is calculated 1 mm above the top tape, similar to a Hall probe experiment. The probe position is more relevant for commercial application than the magnetic field in the tape or between the tapes. The sketch of the modelling case is on figure \ref{comparison.fig} (a).

\begin{figure}[tbp]
\centering
{\includegraphics[trim=0 0 0 0,clip,height=7.0 cm]{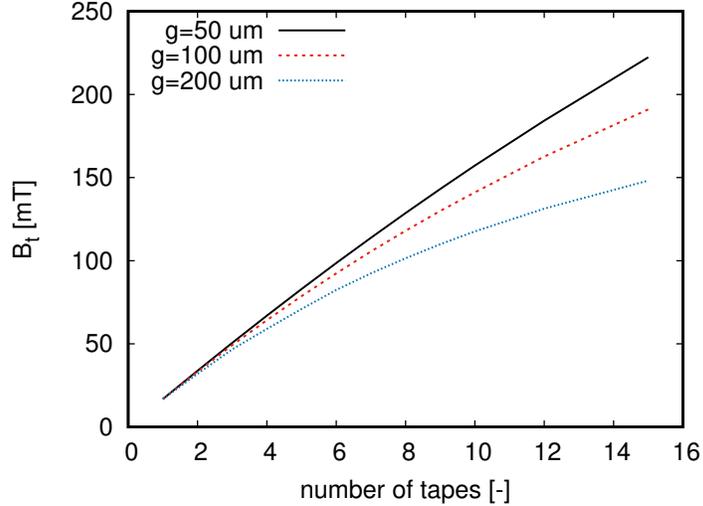}}
\caption{The trapped field increases with the number of the tapes in the stack and reduced gap between superconducting layers. The gap is $g$=50,100,200 $\mu$m.}
\label{trapped_f.fig}
\end{figure}

The trapped field increases with the number of tapes figure \ref{trapped_f.fig}. The big gap of 200 $\mu$m between SC layers causes saturation of the trapped field. The trapped field decreases with the gap $g$, and hence superconducting tapes with thinner Hastelloy, thermal stabilizations\R{, and any additional isolating layers} are more suitable for stack of the tapes based applications. The cause of the decrease in trapped field with increasing the gap is the increase of the overall thickness-to-width ratio, since the contribution of the bottom tapes to the trapped field on the top decreases with the overall thickness. \R{This decay with the thickness is faster in 3D calculations than in cross-sectional 2D. The cause is that in short tapes the magnetic field created by closed loops, being dominated by the dipolar term, decays with the distance $r$ as $1/r^3$, while in 2D the decay is as $1/r^2$.}   

\subsection{Benchmark of MEMEP 3D with FEM}

\begin{figure}[tbp]
\centering
 \subfloat[][]
{\includegraphics[trim=0 0 0 0,clip,height=5.0 cm]{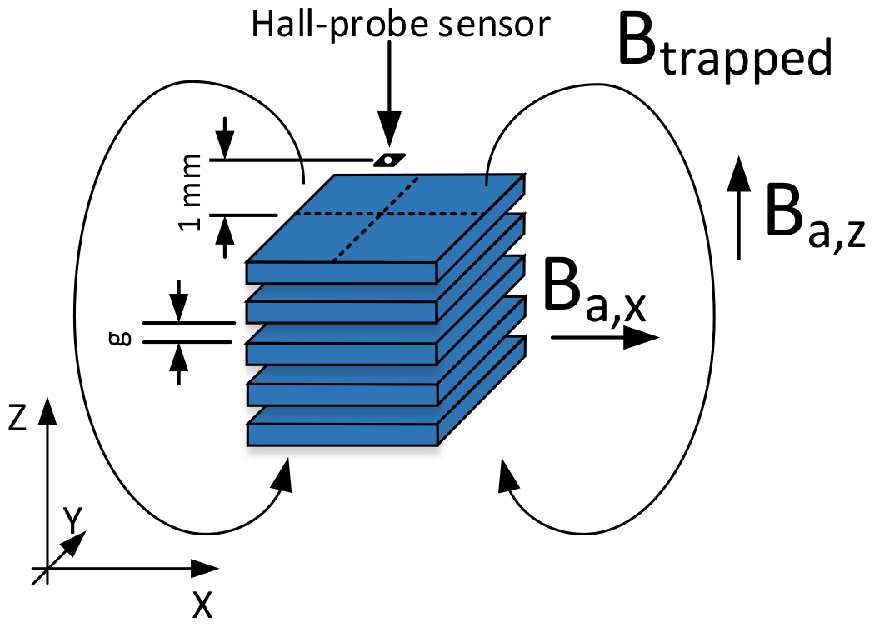}}
 \subfloat[][]
{\includegraphics[trim=0 0 0 0,clip,width=8 cm]{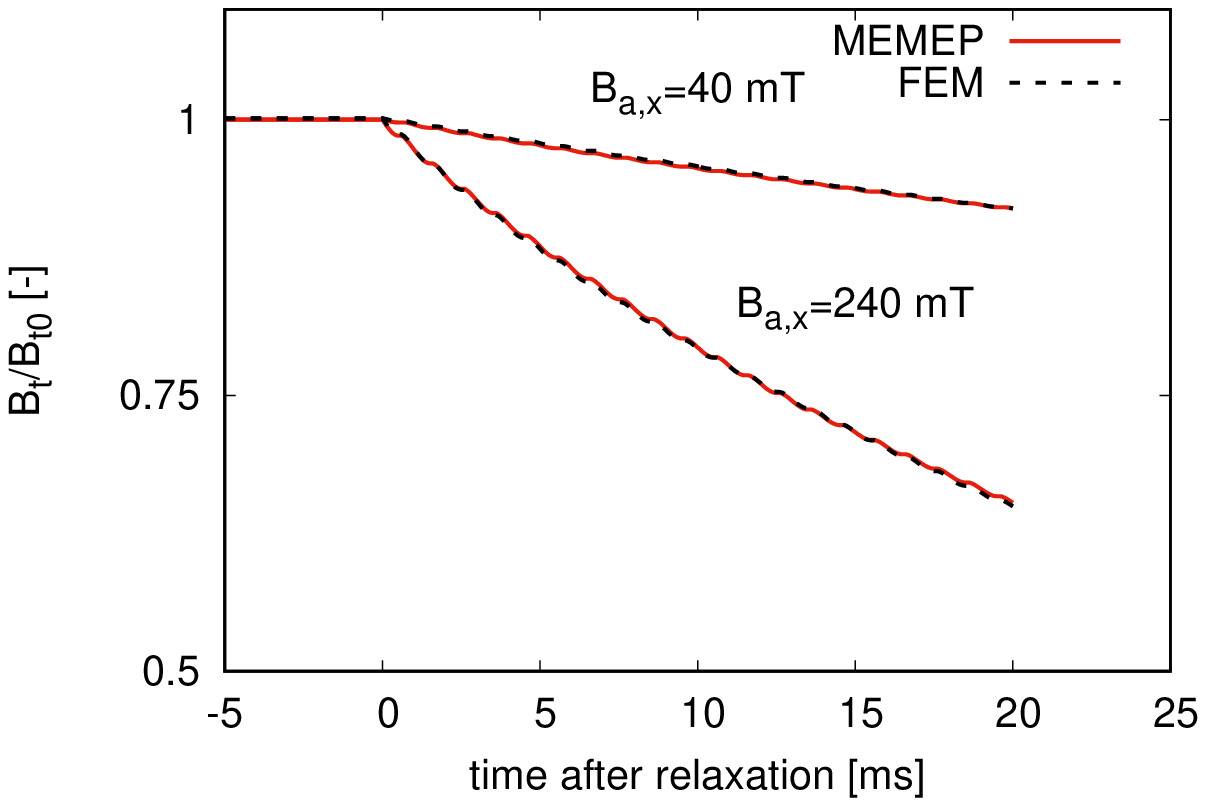}}
\caption{(a) Stack of 5 tapes with Hall probe position and direction of the magnetization and demagnetization fields $B_{a,z}$ and $B_{a,x}$, respectively. The gap between superconducting layers is 200 $\mu m$. (b) The cross-field demagnetization of a two tapes stack calculated by MEMEP 3D and FEM. The methods are in very good agreement for both cross-field amplitudes 40 mT and 240 mT.}
\label{comparison.fig}
\end{figure}

Next, we cross-check the MEMEP 3D method and the Finite Element Method. The comparison case is for a two-tape stack only, for simplicity. The superconducting layer is 10 $\mu$m thin with 200 $\mu$m gap between the layers. The electrical parameters are the same as $J_c$=2.72$\times 10^{9}$ A/m$^2$ and $n$=30. \R{The mesh contains $15\times 15\times 5$ cells per superconducting layer. The computing time for this mesh and the cross-field of 240 mT is 3-4 hours for MEMEP 3D and 14-15 hours for FEM. MEMEP used a computer with i7-4771 CPU at 3.5 GHz and the FEM used workstation with i7-4960X CPU at 3.60 GHz.}

\R{The magnetic moment of the sample at the end of relaxation differs only by 4 \% between both models. Therefore, the screening current calculations are with very good agreement.}
    
The result of the trapped field 1 mm above the top surface is on figure \ref{comparison.fig} (b). The trapped field is normalized by the trapped field at the end of the relaxation time, $B_0$. The value of $B_0$ is 32.4 mT for MEMEP 3D and 28.6 mT for FEM method. The comparison is with good agreement, even though the trapped field is with 11 \% difference \R{This difference is due to inaccuracies in the trapped magnetic field by FEM, due to the relatively coarse mesh in the surrounding air. This could also be the cause of discrepancy in the magnetic moment. Nevertheless, } the demagnetization rate is the same for both methods. This confirms the validity of the MEMEP 3D calculations\R{, with more accuracy than FEM for this configuration}.


\subsection{Cross-field demagnetization for high speed rotating machines (500 Hz)}

The next study is about the entire cross-field demagnetization process in the stack of 5 tapes. The sketch of the modelling case is on figure \ref{comparison.fig} (a) and it uses the same parameters as in the previous sections. The gap $g$ between superconducting layers is 200 $\mu$m and superconducting layer is 10 $\mu$m thick. The time evolution of the magnetizing field, $B_{az}$, and the trapped field on the stack, $B_t$, are on figure \ref{wave_form.fig}. At time 1000 s, the sinusoidal ripple field, $B_{ax}$, of 500 Hz is switched on.

\begin{figure}[tbp]
\centering
{\includegraphics[trim=0 0 0 0,clip,height=4.5 cm]{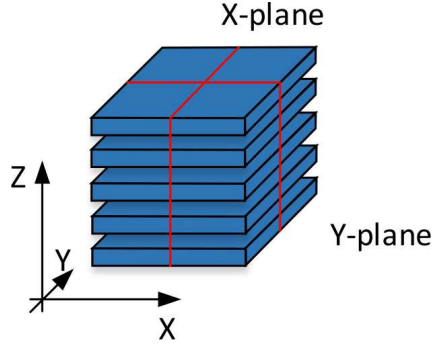}}
\caption{The cross-sectional planes are at $x=6$ mm and $y=6$ mm for the current density colour maps.}
\label{sketch2.fig}
\end{figure}

\begin{figure}[tbp]
\centering
{\includegraphics[trim=0 0 0 0,clip,height=5.0 cm]{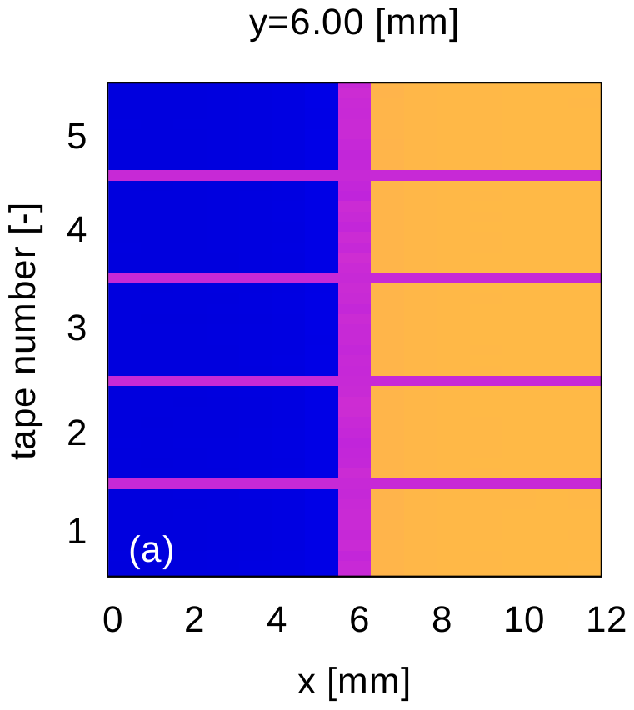}} 
{\includegraphics[trim=20 0 0 0,clip,height=5.0 cm]{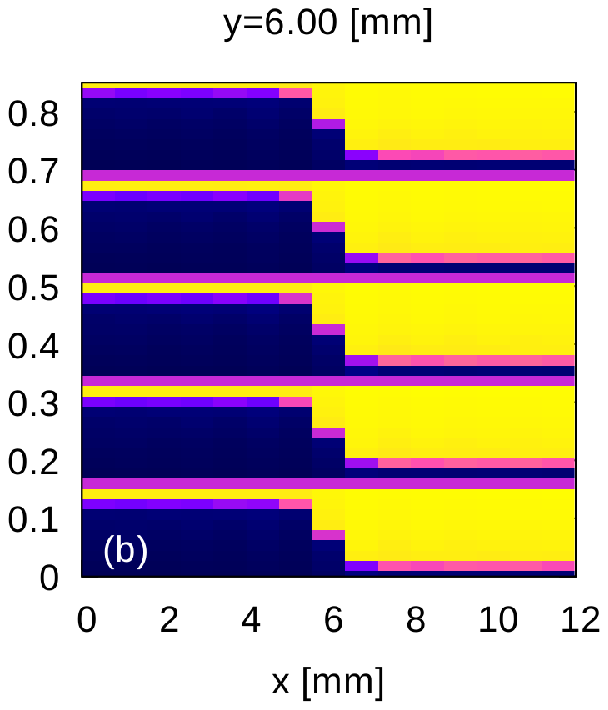}} 
{\includegraphics[trim=45 0 0 0,clip,height=5.0 cm]{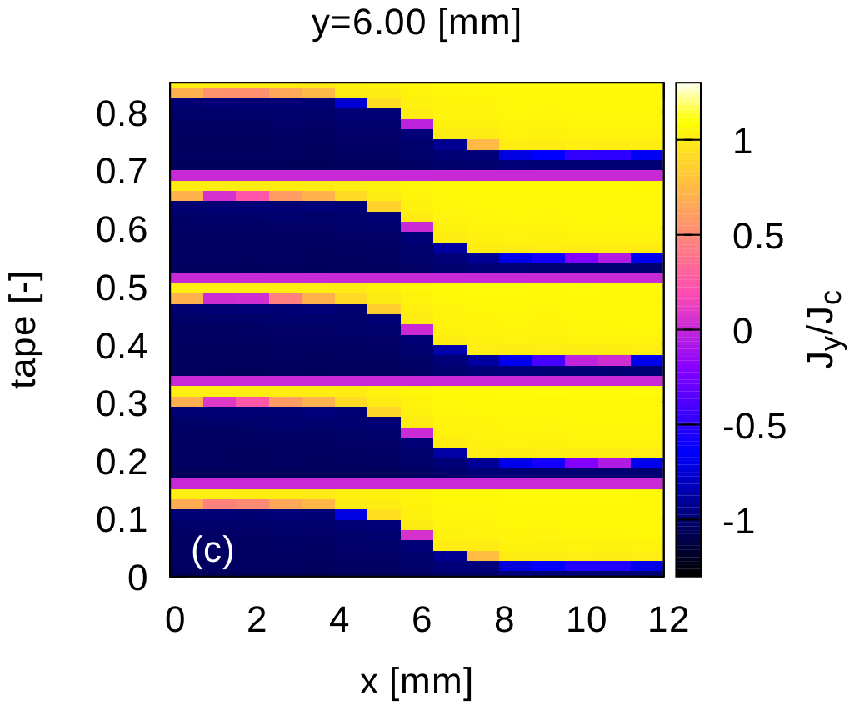}} \\
\vspace{1 mm}
{\includegraphics[trim=0 0 0 0,clip,height=5.0 cm]{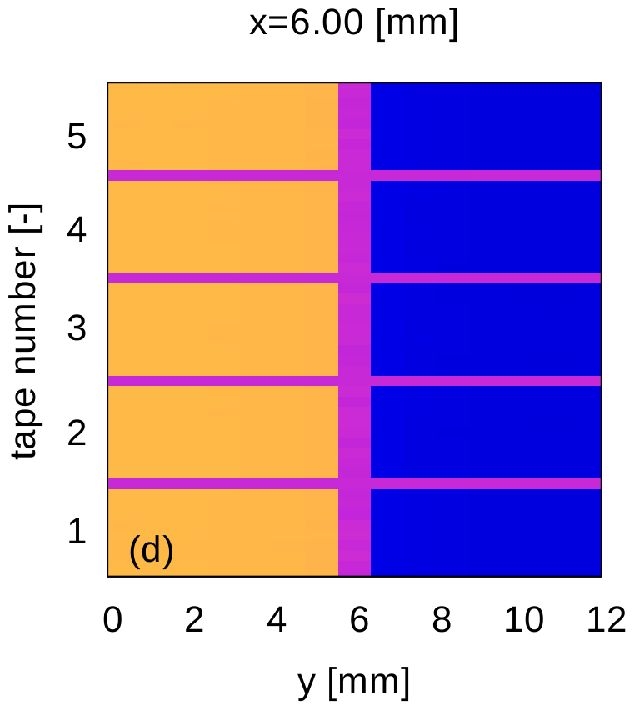}} 
{\includegraphics[trim=10 0 0 0,clip,height=5.0 cm]{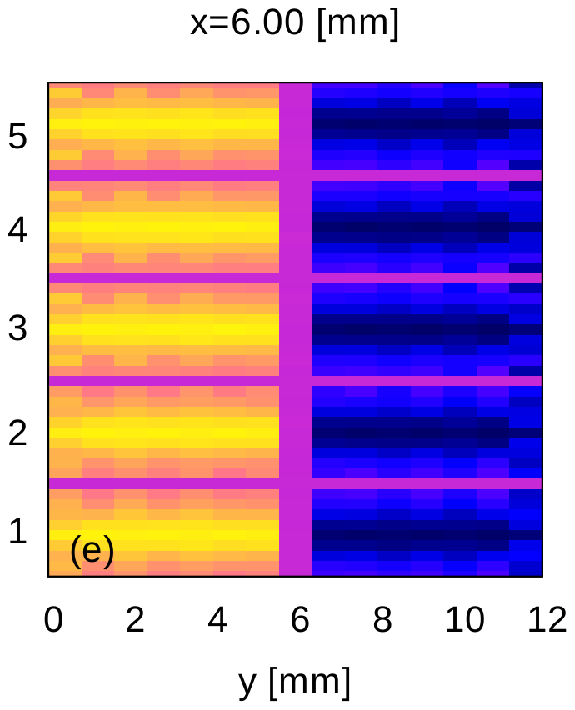}} 
{\includegraphics[trim=10 0 0 0,clip,height=5.0 cm]{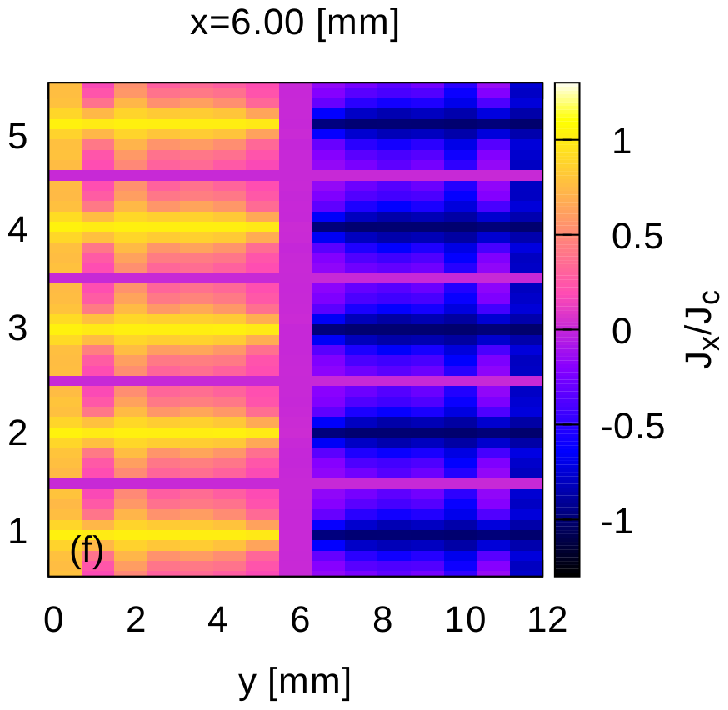}}
\caption{3D current density colour maps during cross-field demagnetization. The perpendicular component of the current density in the cross-sections at the middle of the stack at $x$=6 mm, $y$=6 mm, (a, d) after 15 minutes relaxation (b, e) at first positive peak of ripple field 240 mT (c,f) at last positive peak	(10th cycle). The air and the superconducting cells are plotted on the colour maps with the same high for better visibility. However, the calculation uses the real dimensions of the sample.}
\label{current.fig}
\end{figure}

\begin{figure}[tbp]
\centering
{\includegraphics[trim=0 0 0 0,clip,height=6.5 cm]{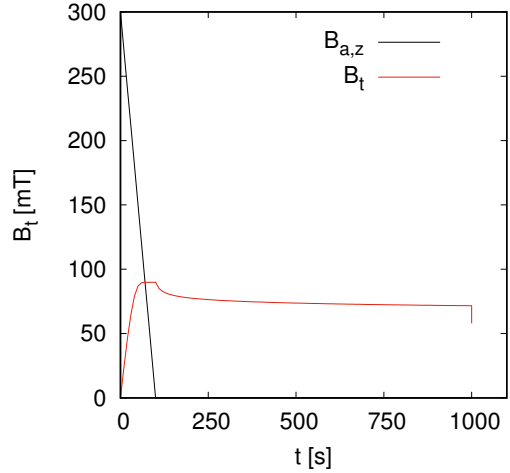}}
\caption{Time evolution of the \R{magnetizing} applied field, $B_{a,z}$, and the trapped field, $B_t$, on the stack. \R{Field cooling magnetization and subsequent relaxation end at times 100 s and 1000 s, respectively. Afterwards, the stack experiences cross-field demagnetization.}}
\label{wave_form.fig}
\end{figure}

The study of the current density inside the stack is on the $x$ and $y$ cross-section planes, as defined in figure \ref{sketch2.fig}. The current density maps (figure \ref{current.fig}) are in the real scale except the $z$ coordinate. Since the tapes are 10 $\mu$m thick and the gap is 200 $\mu$m wide, the variations in the thickness of the current density are not visible in the real scale. Therefore, the maps contains the real data, but superconducting cells are shown with the same high as the the gap cells. Since we use an odd number of cells in the $x$ and $y$ directions, there appears a central cell with zero current density in figure \ref{current.fig}(a,d-f), causing the vertical purple line. The study case is with cross-field amplitude 240 mT. 

The stack at the end of relaxation is fully saturated with $J_x\approx J_c$ and $J_y\approx J_c$ on figures \ref{current.fig} (a) and \ref{current.fig} (d), respectively. The $J_z$ component is very small, around 0.001$J_c$ due to the thin film shape. The current density magnitude is below the critical current density $J_c$, because of flux relaxation. 

The applied cross-field at the first positive peak causes the small penetration of the screening current from top and bottom of each individual tape as seen in figure \ref{current.fig} (b). The penetration front on the $y$ plane rewrites the remanent state of the current density $J_y$ to the positive sign from the top and with negative sign from the bottom of each tape (figure \ref{current.fig} (b)). This process can be explained by the Bean model of the infinite thin strip \cite{Brandt02PRL} or other numerical 2D modelling \cite{Baghdadi18SR,Campbell17SST}. In our stack, the penetration front contains both the $J_y$ and $J_z$ components, even though $J_z$ is very small. The $x$ plane section of figures \ref{current.fig} (d-f) shows that the $J_x$ current density from remanent state is progressively decreasing at each cycle by the effect of $J_y$ (and also $J_z$ in a lesser amount) \R{caused by the ripple field, which penetrates from} the top and bottom of each individual tape. \R{This is the same qualitative behavior found for superconducting bulks \cite{Kapolka18IES}. The oscillations in the current profiles in figure \ref{current.fig}(e,f) are due to numerical error.}

Finally, we study in detail the current density maps at the tenth positive peak of the cross-field. At the $y$ plane, the $J_y$ current density component penetrates slightly more than after the end of the 1st cycle. The penetration depth is in 2 cells from the top and bottom of each layer within the total of 9 cells per superconducting layer (figure \ref{current.fig} (c)). The lowest penetration is in the inner tapes. The cause is that the inductive coupling with the rest of the tapes is the largest. This enhanced inductive coupling slows down the demagnetization decay due to the dynamic magneto-resistance \cite{Brandt04PhC,Brandt02PRL}, being the decay exponential for large enough time. At the $x$-plane, the $J_x$ component is almost completely concentrated in 3 cells in the centre of each superconducting layer (figure \ref{current.fig} (f)). The current maps show slow penetration of the current front and erasing process of the remnant state with increasing the number of cycles. Although $J_z$ plays a role in the cross-field demagnetization, its value is much below $J_c$, and hence reducing $J_c$ in the $z$ direction will not cause any significant change.

\begin{figure}[tbp]
\centering
{\includegraphics[trim=0 0 0 0,clip,width=10 cm]{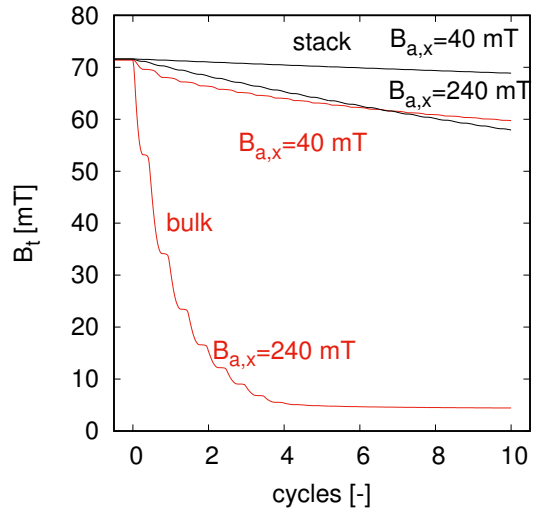}}
\caption{The cross-field demagnetization \R{of the trapped field, $B_t$} of the \R{studied} stack of tapes \R{demagnetizes slower than its equivalent bulk (sketch in figure \ref{sketch4.fig}). Calculations for} transverse field amplitudes of 40 and 240 mT \R{with} 500 Hz \R{frequency}.}
\label{bulk.fig}
\end{figure}

The trapped magnetic field $B_t$ during the whole process is calculated 1 mm above the top surface (figure \ref{wave_form.fig}). After relaxation, we applied cross-fields with two different amplitudes, 40 and 240 mT, in order to see the behaviour in the fields below and above the penetration field $B_p$ of the stack. The parallel penetration field $B_{p,\parallel}$ of 1 tape is 17 mT according to the slab Critical-State model \cite{bean64RMP,navau13IES}, 
\begin{equation}
\ B_{p,\parallel}=\mu_0J_cd/2.
\label{T_J}
\end{equation}
The demagnetization rate for 40 mT cross-field is very low, with 3.9$\%$ drop of the trapped field after 10 cycles (figure \ref{bulk.fig}). The higher cross-field of 240 mT makes 19.1 $\%$ reduction of the trapped field. The roughly linear demagnetization is consistent with Brandt's predictions, where there is linear decay for the first few cycles \cite{Brandt02PRL}. However, the method of \cite{Brandt02PRL} is based on Bean model and for a single tape. For applied fields above the penetration field, as is the case of both 40 and 240 mT, demagnetization will continue until the entire sample is demagnetized.

\R{For high demagnetization rates, such as high ripple field amplitudes or thick superconducting layers, we observe steps in the demagnetization of the time evolution (figures \ref{thickness.fig} and \ref{bulk.fig}). These steps are also present in bulks (see figure \ref{bulk.fig} and \cite{Kapolka18IES}) and are due to the changes in the rate of current density penetration within an AC cycle. Current penetration from the top and bottom layers of each tape slows down at the AC peaks, causing a plateau in the trapped field.}

\begin{figure}[tbp]
\centering
 \subfloat[][]
{\includegraphics[trim=0 0 0 0,clip,height=4.5 cm]{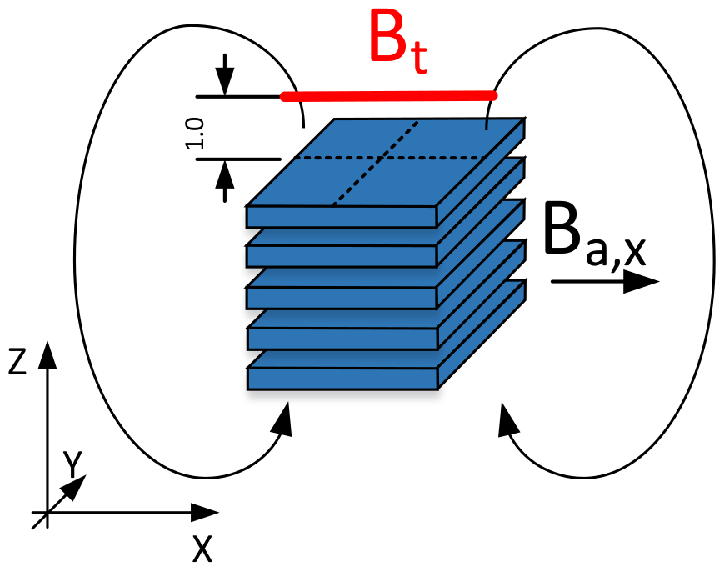}}
 \subfloat[][]
{\includegraphics[trim=0 0 0 0,clip,height=5.5 cm]{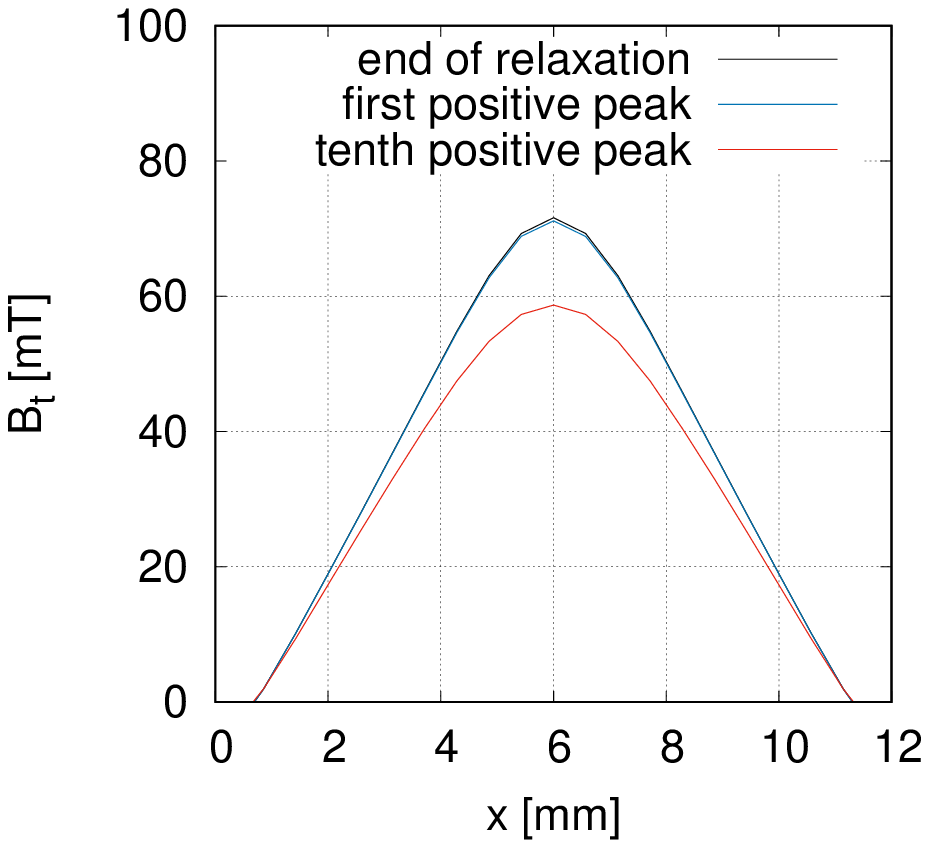}}
\caption{(a) The position of calculated trapped field profile $B_t$ is 1 mm above the stack. (b) The trapped field profile decreases during demagnetization by $B_{a,x}$ with amplitude 240 mT and 500 Hz.}
\label{profile.fig}
\end{figure}

A more detailed trapped field profile is calculated along the (red) line $B_t$ above the sample on figure \ref{profile.fig} (a). The profile has the usual symmetric peak at the end of the relaxation time (figure \ref{profile.fig} (b) black curve). The first positive peak of the cross-field of 240 mT makes a small reduction of the trapped field [figure \ref{profile.fig} (b) blue curve], since the current density has not changed significantly due to transverse field. The last positive peak of the cross-field is shown on figure \ref{profile.fig} (b) as red line. The trapped field peak is always symmetric without any shift, contrary to cubic bulk samples \cite{Kapolka18IES}. The cause of this difference is the thin film shape of tapes, as follows. For the bulk, the trapped field depends on the current distribution across the thickness, with a higher contribution for $J$ closer to the top surface. Since $J$ does not have mirror symmetry towards the $yz$ plane, (only inversion symmetry towards the bulk center) the trapped field on the surface is not symmetric. In contrast, the trapped field in the thin films only depends on the average $J$ across the tape thickness, being variations on this dimension irrelevant. Since the thickness-average $J$ does have mirror symmetry with respect to the $yz$ plane for each tape of the stack, the trapped field on the surface also presents this mirror symmetry.

\subsection{Comparison of cross-field demagnetization in a stack of tapes and bulk}

\begin{figure}[tbp]
\centering
 \subfloat[][]
{\includegraphics[trim=0 0 -20 0,clip,width=5.7 cm]{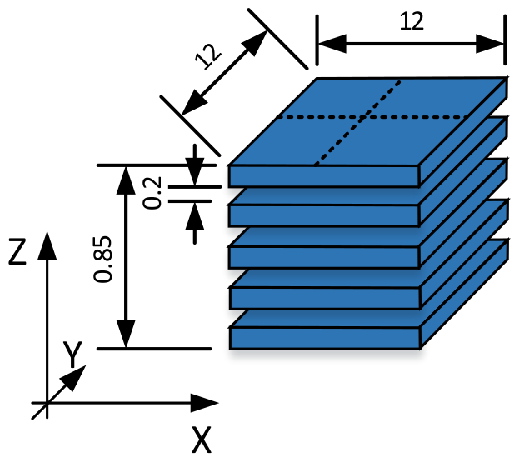}}
 \subfloat[][]
{\includegraphics[trim=0 0 0 0,clip,width=5 cm]{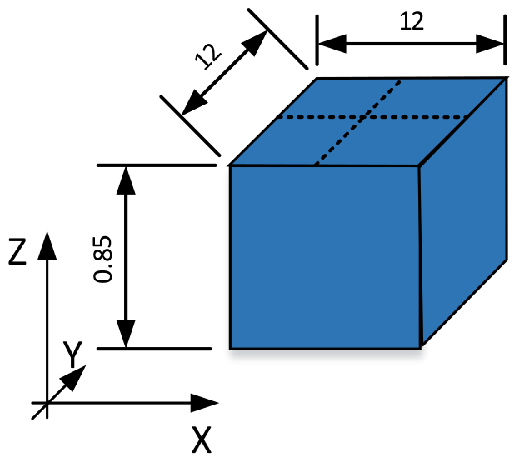}} 
\caption{Bulk and stack of tapes with the same geometrical dimensions and engineering critical current density for comparison. Dimensions are in mm.}
\label{sketch4.fig}
\end{figure}

There are two alternatives for high temperature super-magnets: stacks of tapes and bulks. Both candidates broke the world record of trapped field, being above 17 T with slightly higher values for the stack \cite{Durrell14SST,Patel18SST}. However, both behave differently under cross-fields. Therefore, we performed a short simple comparison between them. We used the same geometry for 5 tapes stack as it was mentioned above. We calculated the engineering current density for the stack $J_{ce}$=160 MA/m$^2$ and set it as critical current density $J_c$ for the bulk. The samples with the size dimensions are in figure \ref{sketch4.fig}. We estimated the parallel penetration field of the equivalent bulk from the slab approximation       
\begin{equation}
\ B_{p,\parallel}\approx\mu_0J_{ce}d_{all}/2,
\label{T_J}
\end{equation}
being $d_{all}$=0.85 mm the overall stack thickness and $B_{p,\parallel,bulk}\approx$85.4 mT.

The trapped field 1 mm above the top surface at the end of the relaxation time of 900 s is similar for the stack (71.6 mT) and the bulk (71.4 mT), because of similar parameters. The end of the relaxation time is marked as 0 cycle on figure \ref{bulk.fig}. The bulk shows significant trapped field drop, around 93$\%$ in the first four cycles of 240 mT cross-field, being larger than the bulk penetration field (85 mT). In the case of low cross-field amplitude of 40 mT, the drop is around 16 $\%$. For large number of cycles, the trapped field will decrease until the quasi-stable state is reached, because the ripple field amplitude is below the parallel penetration field of the bulk \cite{Brandt02PRL}, the latter being dominated by a slow flux creep decay \cite{Srpcic19SST}. The stack shows much slower demagnetization rate for the high cross-field of 240 mT and the trapped field drop in 10 cycles is only 19 $\%$. Nevertheless, demagnetization should continue until it completely demagnetizes the sample, because the ripple field is above the parallel penetration field of one tape (17mT). The same behaviour is observed for the low cross-field of 40 mT with very low trapped field reduction of 3.9$\%$ at 10 cycles. Therefore, for the high cross-field cases (ripple field above the parallel penetration field of the bulk) the stack of tapes is more suitable. However, in the case of low fields the bulk is more suitable because the asymptotic trapped field does not vanish after many cycles. Nevertheless, if the super-magnet is submitted to relatively low number of cycles, the stack of tapes are preferred in any case. Applications with low frequency ripples and built-in re-magnetization, such as certain low-speed motors and wind turbines, might also favour stacks of tapes, because of less re-magnetization.       


\subsection{Cross-field demagnetization: measurements and modelling}

The last study is about measurements and comparison with calculations. The stack consists of 5 tapes and the parameters are in the table \ref{input.tab}. The details about the sample and the measurements are given in section \ref{s.methodology}. The measurements are performed for two cross-field frequencies: 1 and 10 Hz. 

\begin{figure}[tbp]
\centering
{\includegraphics[trim=30 0 30 0,clip,width=7 cm]{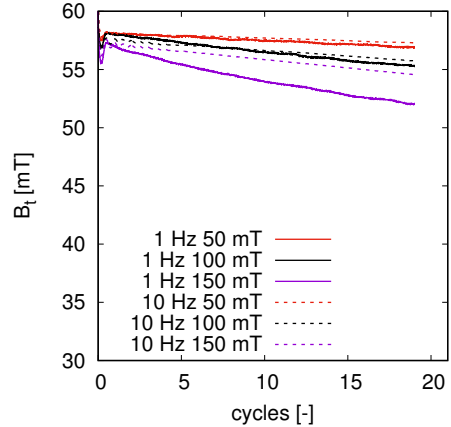}}
\caption{The \R{measured} demagnetization rate increases with the cross-field amplitude. There is \R{also a} low frequency dependence\R{, although} the demagnetization rate is similar for both frequencies.}
\label{measurements.fig}
\end{figure}

The demagnetization rate increases with the field (figure \ref{measurements.fig}). Rate per cycle at low frequencies depends on the frequency only slightly \cite{Baskys18SST}. The reason is that the higher frequency of the applied field causes higher electric field, and hence the current density increases. This reduces both the penetration field and the demagnetization rate \cite{Kapolka18IES,Sander10JCS,Thakur11SSTa,Thakur11SST,acreview}. \R{The measurements also show increased frequency dependence with field (figure \ref{measurements.fig}). A possible cause might be the decrease of the power-law exponent with the applied field amplitude [figure \ref{nB.fig}(a)]. The oscillations observed in the measurements, are more likely due to experimental noise.} 

\begin{table}[!t]
\renewcommand{\arraystretch}{1.3}
\caption{Input parameters of the measurements and calculation.}
\label{input.tab}
\centering
\begin{tabular}{|c||c|}
\hline
Size [mm] & 12x12x0.0015\\
\hline
${J_{c,self}}$ [A/m${^2}$] & 2.38${\times 10^{10}}$\\
\hline
${B_{az,max}}$ [T] & 1.0\\
\hline
Ramp rate [mT/s] & 10\\
\hline
Relaxation [s] & 300\\
\hline
${E_c}$ [V/m] & 1e-4\\
\hline
${f_{ax}}$ [Hz] & 0.1,1\\
\hline
${B_{ax}}$ [mT] & 50,100,150\\
\hline
n [-] & 30\\
\hline
\end{tabular}
\end{table}

\R{For the model, we take three different assumptions for $J_c$: constant $J_c$, isotropic $J_c(B)$ with $B$ being the local modulus of the magnetic field, and anisotropic $J_c(B,\theta)$. Only the latest approach provides realistic predictions. The model uses 1.5 $\mu$m thin superconducting layer with 220 $\mu$m gap between them, being the rest of parameters the same as for the measurements (table \ref{input.tab}).}

\begin{figure}[tbp]
\centering
{\includegraphics[trim=0 0 0 0,clip,width=12 cm]{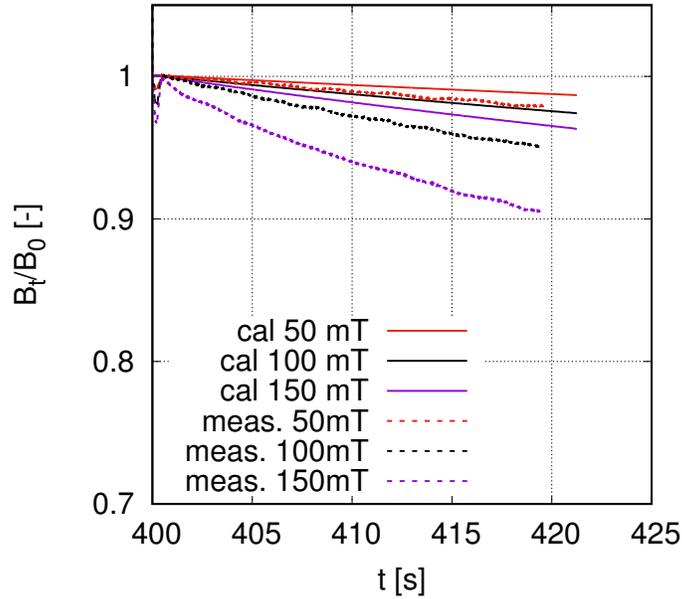}}
\caption{Comparison of the measurements and calculations (constant $J_c$ = 1.85$\times 10^{10}$ A/m$^2$) of 1 Hz cross-field demagnetization. The constant $J_c$ underestimates the demagnetization rate due to missing $J_c(B)$ dependence.}
\label{Jc.fig}
\end{figure}

\R{The coarsest predictions are for constant $J_c$} (figure \ref{Jc.fig}). The \R{measured} $I_c$ of the \R{12 mm wide} SuperOx tape is \r{450 A} at self-field (or \R{$J_c$=2.5$\times 10^{10}$ A/m${^2}$}). We reduced $J_c$ by \R{26~\%} to the value $J_c$= 1.85${\times 10^{10}}$ A/m$^2$, in order to get similar trapped field, 58.9 mT, at the end of the relaxation time as the measurements, 58.1 mT. \R{Naturally, this is an artificial correction of the stack self-field effect, and hence for constant $J_c$ the predicting power regards only to the demagnetization rate.} \R{However, the predicted} demagnetization rate is \R{substantially} lower than the measured one. The reason is that the assumed critical current density is too high, because of the missing $J_c(B,\theta)$ dependence in the model. The magnetic field reduces the critical current density, and hence it increases the demagnetization rate.

\begin{figure}[tbp]
\centering
{\includegraphics[trim=0 0 0 0,clip,width=12 cm]{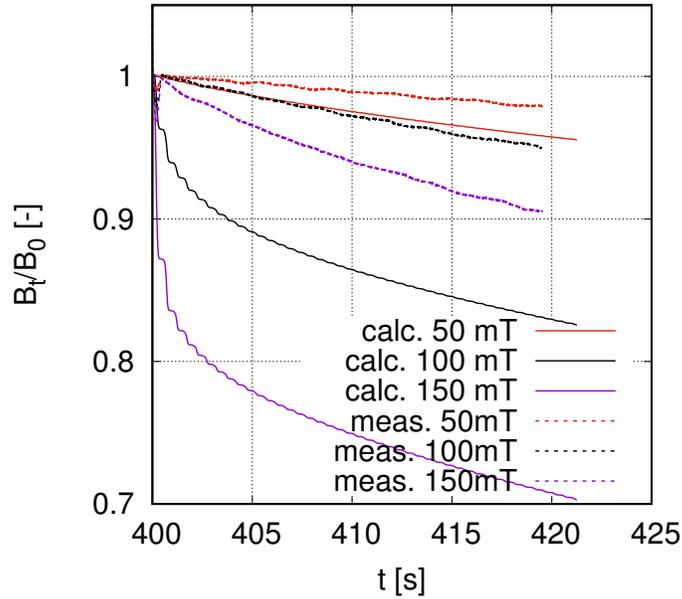}}
\caption{Comparison of the measurements and calculations with $J_c(B)$ measured data of 1 Hz cross-field. The $J_c(B)$ dependence overestimates the demagnetization rate due to missing anisotropy dependence.}
\label{JcB.fig}
\end{figure}

Another comparison between the model with $J_c(B)$ dependence and measurements is on the figure \ref{JcB.fig}. The $J_c(B)$ data was measured on the 4 mm wide SuperOx tape (figure \ref{JcB_data.fig}). The critical current per tape width at self-field for the measured tape (37.8 A/mm) is roughly the same as the 12 mm wide tape used in the stack (35.8 A/mm), being the latter value the minimum stated one by the producer. The theoretical difference is 5$\%$, which is very small. The average tape $I_c$ could be higher, around 440 A or 450 A regarding typical deviations in SuperOx tapes, and hence even more close to that in the calculations. By now, we assume an isotropic $J_c(B)$ dependence, taking the measured $J_c(B,\theta)$ values at perpendicular applied field. Then, we assume an isotropic angular dependence in the model. 
The cross-field is parallel to the tape surface, and hence the actual critical current density is larger than that in the model, also presenting lower reduction under magnetic fields than assumed. This is the reason why the demagnetization rate is overestimated for the high cross-fields of 100 mT and 150 mT.

\begin{figure}[tbp]
\centering
{\includegraphics[trim=0 0 0 0,clip,width=12 cm]{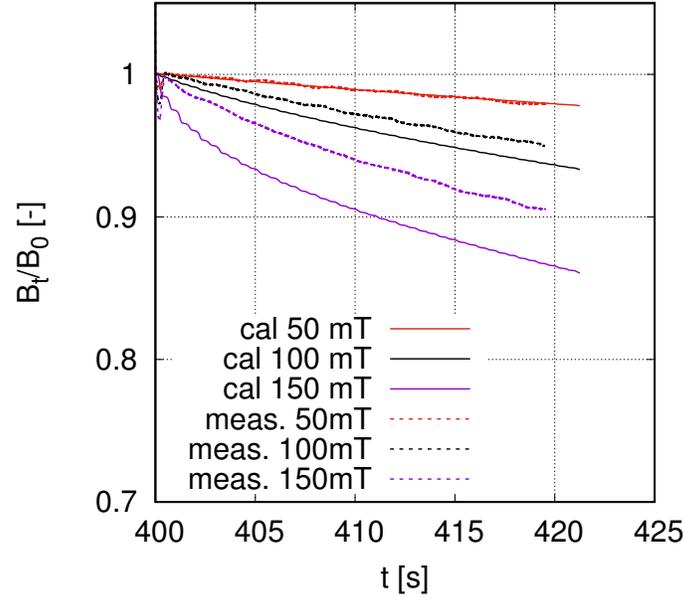}}
\caption{The comparison of measurements and calculations with $J_c(B,\theta)$ measured data of 1 Hz transverse field. The calculation agrees very well for low cross-field up to 50 mT. The cross-field above 50 mT requires $J_c(B,\theta)$ measured data with a parallel component to the current path. However, the results for higher fields are with good accuracy, around 4$\%$.}
\label{Jcbt.fig}
\end{figure}

\begin{figure}[tbp]
\centering
\subfloat[][]
{\includegraphics[trim=40 0 50 0,clip,width=7 cm]{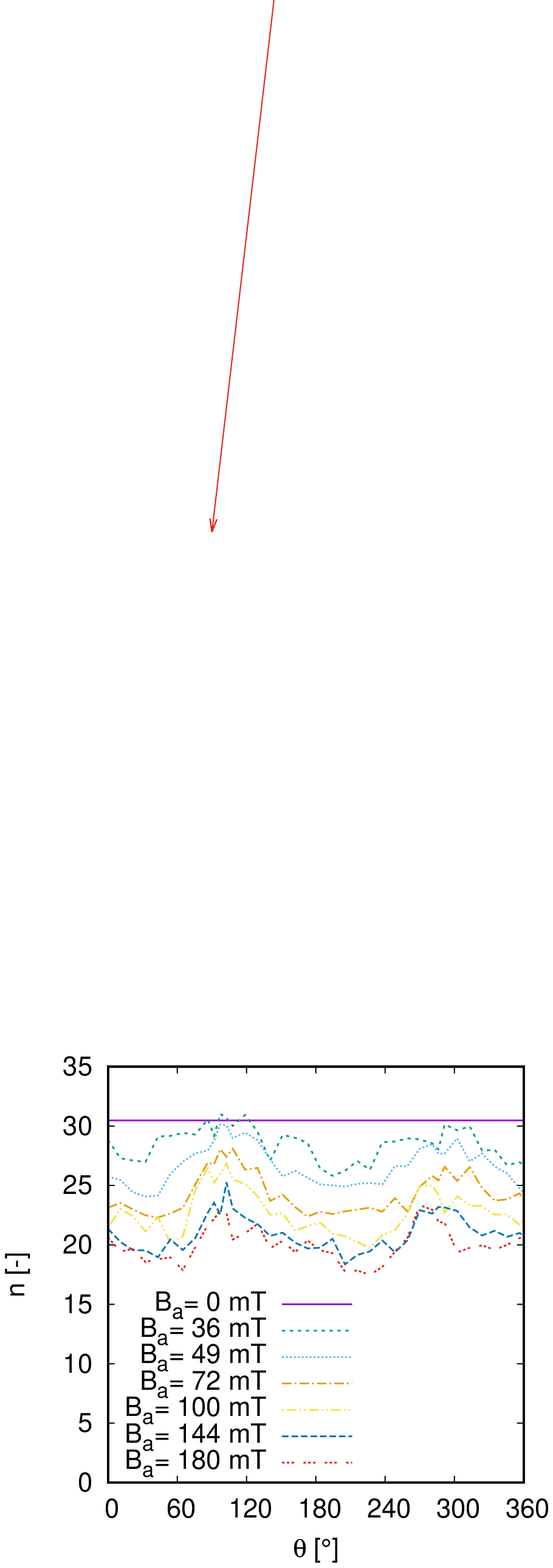}}
\subfloat[][]
{\includegraphics[trim=40 0 50 0,clip,width=7 cm]{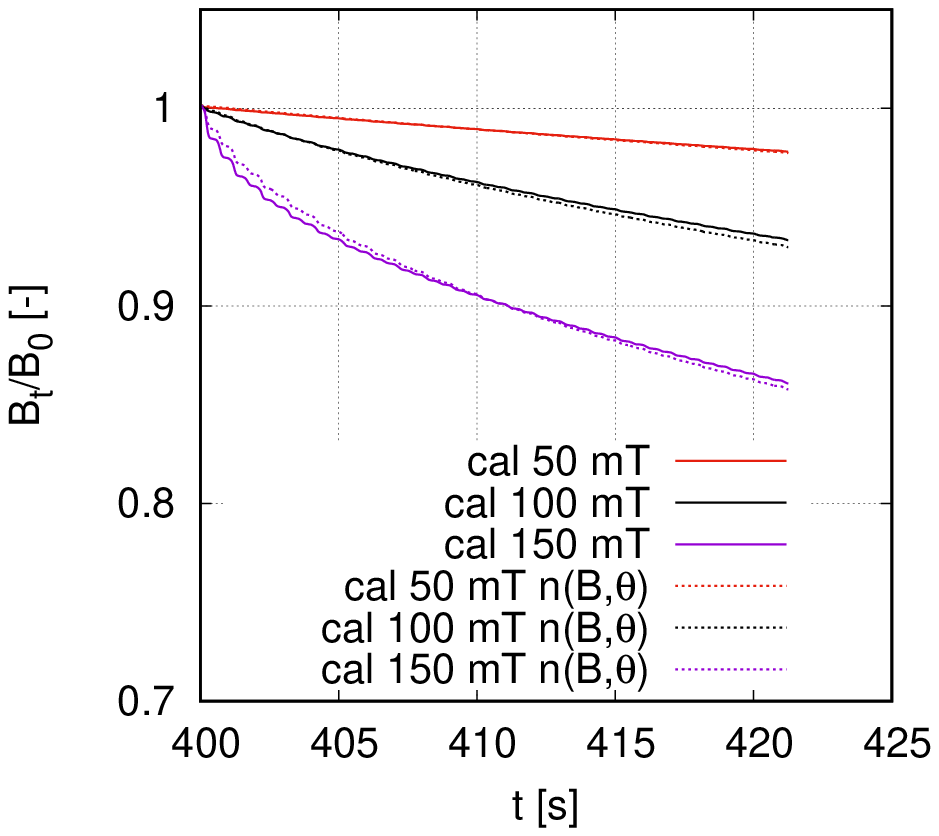}}
\caption{(a) The $n(B,\theta)$ measured data on a 4 mm wide SuperOx tape\R{, measured in Bratislava by the set-up in \cite{seiler19SST}}. (b) The comparison of calculation with constant $n$=30 and $n(B,\theta)$ dependence, both cases use $J_c(B,\theta)$ dependence. Using $n(B,\theta)$ slightly reduces the demagnetization rate for a few number of cycles, but later on it is increased slightly.}
\label{nB.fig}
\end{figure}

The last comparison here uses the measured $J_c(B,\theta)$ dependence. Now, the model agrees very well with measurements (figure \ref{Jcbt.fig}). The lowest deviation \R{of the trapped field} at the last cycle is 0.2 \% is for 50 mT and 1.2 \% for 100 mT cross-field. The \R{highest} cross-field \R{amplitude causes the highest} demagnetization in the calculation, being the difference 4.0 \%. \R{For low demagnetization, the demagnetization rate, $\beta$, is more relevant than the trapped field, which we can define as 
\begin{equation}
\beta \equiv [B_t(t_{\rm i})-B_t(t_{\rm f})]/(t_{\rm f}-t_{\rm i})   \label{demrate}
\end{equation}
with $t_{\rm i}$ and $t_{\rm f}$ being the time at the beginning and end of the demagnetization process, respectively. For our configuration, which is similar to that in other works \cite{Baghdadi14APL,Campbell17SST,Liang17SST,Baghdadi18SR}, the error regarding this quantity is more strict than for the trapped field (table \ref{error.tab}), resulting in errors below 20 \% for ripple fields of 100 mT or below.}

We also study the effect of the measured $n(B,\theta)$ dependence (figure \ref{nB.fig} (a)) compared to the constant $n=30$ assumption. The demagnetization rate is slightly changed with $n(B,\theta)$ dependence (figure \ref{nB.fig} (b)). The local $J$ increases with decreasing the $n$ value, and hence it changes the demagnetization rate. There is a slight reduction of the demagnetization rate at the first few cross-field cycles. However, later on the demagnetization rate overlaps with the constant $n$ curve and slightly increased. The demagnetization rate is more influenced for cross-field above 50 mT.

\begin{table}[!t]
\renewcommand{\arraystretch}{1.3}
\caption{\R{Deviation of the MEMEP 3D modelling results of trapped field, $B_t$, and demagnetization rate defined as $[B_t(t_{\rm i})-B_t(t_{\rm f})]/(t_{\rm f}-t_{\rm i})$, being with $t_{\rm i}$ and $t_{\rm f}$ the time at the beginning and end of the demagnetization process, respectively. The second error estimation is more strict for low demagnetization rates. Results are for the 20th cycle of the configuration in figure \ref{Jcbt.fig}.
}}
\label{error.tab}
\centering
\begin{tabular}{llll}
\hline
\hline
{\bf Ripple field amplitude [mT]} & \bf 50 & \bf 100 & \bf 150\\
\hline
Deviation of trapped field [\%] & 0.2 & 1.2 & 4.0\\
Deviation of demagnetization rate [\%] & 3.0 & 19 & 29\\
\hline
\hline
\end{tabular}
\end{table}

There are several reasons for reduction of the accuracy in the model. The $J_c(B,\theta)$ data covers correctly only the $J_c$ in the $y$-plane position (figure \ref{sketch2.fig}), where the current is perpendicular to the magnetic field. However, at the $x$-plane, the current density presents a large component parallel to the applied magnetic field, being in the $x$ direction. This is the so-called force-free configuration \cite{Kapolka19SST}, where $J_c$ should be different than the typical $J_c(B,\theta)$ measurements with $\vB$ always perpendicular to the transport direction. Since \R{measurements show that} $J_c$ in force-free configuration is often higher \R{\cite{Clem11SST,Vlasko15FL}}, this could explain the overestimated cross-field demagnetization in the model. Measurements of \R{solid-angle} dependence, $J_c(B,\theta,\phi)$ with $\phi$ being the angle of $\vB$ with the current density, are scarce for any type of sample \cite{Clem11SST} and missing for this particular tape. The cause is the complexity of the measurements, requiring a double goniometer \cite{Herzog94RSI,Durrell03PRL,Withnell09IES,Lao17SST}. The model uses $J_c(B,\theta)$ data for both components $J_x$ and $J_y$, and hence there is discrepancy between the model and the real measurements in the highest cross-fields. \R{The discrepancy between measurements and calculations could also be due to variations in the magnetic-field dependence of $J_c$ between the measured one and the SuperOx tapes in the stack.}

\subsection{\R{Error caused by 2D modeling assumptions}}

\begin{figure}[tbp]
\centering
{\includegraphics[trim=45 0 50 0,clip,width=7 cm]{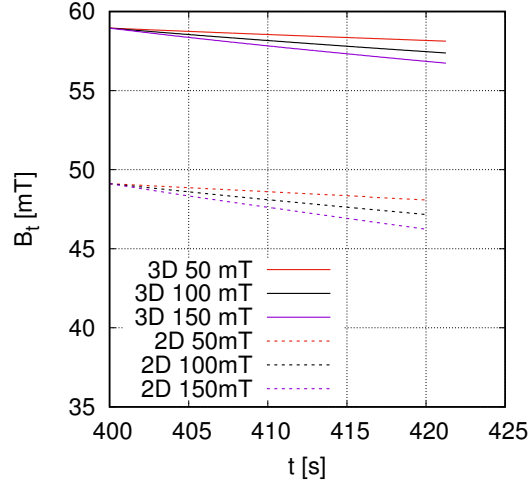}}
\caption{\R{Using the infinitely long assumption (2D modeling) incurs to around 20 \% error in the trapped feild, $B_t$, and a higher demagnetization rate. 2D modeling also disables the possibility to take force-free effects into account, as well as to study the parallel component of $\vJ$ to the ripple field.}}
\label{2D_3D.fig}
\end{figure}

\R{
Finally, we check the relevance of 3D modeling, in contrast to cross-sectional 2D computations. Since this is entirely a geometrical effect, the constant $J_c$ assumption is sufficient. For the calculations, we use the same conditions as the previous section, about comparison to experiments.

As seen in figure \ref{2D_3D.fig}, the trapped field for 2D modeling is around 20 \% lower at the end of relaxation (time 400 s). In addition, there are also substantial differences on the relative demagnetization rate, as defined in equation (\ref{demrate}). 2D modeling predicts {\bf faster} demagnetization rate by 33, 35, and 39 \% for 50, 100, and 150 mT ripple field amplitudes, respectively, as calculated for the 20th cycle. This will worsen the agreement with experiments, where the 3D model already over-estimates the demagnetization rate for 100 and 150 mT, being possibly caused by neglecting force-free effects. Since 2D modeling also inherently neglects force-free effects, geometrical and force-free errors accumulate, which might cause a deviation as large as 60 \% for the 20th cycle and high ripple fields.

In addition, 2D modelling cannot study the parallel component of $\vJ$ to the ripple field. This component, $x$ component in figures \ref{sketch2.fig} and \ref{current.fig}, has roughly the same contribution to the trapped field as the $y$ component, being the only one computed in 2D modeling. 2D modeling also disables the possibility to take force-free effects into account, potentially incurring to further errors.

}

\section{Conclusions}

\R{This article presents 3D modeling of cross-field demagnetization in stacks of REBCO tapes for the first time, being this a substantial achievement according to the methodology in \cite{stenvall19IES}. This 3D modeling enables to take the sample end effects into account, being essential in laboratory experiments, usually made of square tapes, and certain applications, such as axial flux motors or Nuclear Magnetic Resonance (NMR) magnets. The 3D geometry and the used MEMEP 3D model also enables to take force-free effects (or also $ab$ axis anisotropy) into account, being a possible cause of disagreement with experiments at high ripple fields.}

\R{The studied shape, modeled by the MEMEP 3D method, is very challenging, since we need to keep the real superconductor thickness, of the order of 1 $\mu$m, and mesh the layer across its thickness, achieving cells of the range of 100 nm. This results in an aspect ratio of the 3D cells of around 5000.} The MEMEP 3D modelling tool showed the full cross-field demagnetization process in a 5 REBCO tape stack and the 3D screening current path. \R{This enables to study also the current component parallel to the ripple field; in contrast to 2D modeling, which can describe the perpendicular component only}. 

\R{In this article, we present a complete study of cross-field demagnetization by 3D modeling. Among other issues, we confirm the need of the real thickness of the tape, around 1 $\mu$m, by 3D modeling up to 10 cycles. When comparing the stack with the equivalent bulk, the stack demagnetizes slower.} However, we expect that the bulk reaches an stable state without further drop of the trapped field under cross-fields lower than the \R{bulk} penetration field.

The measurements of the 5 tapes stack assembled from 12 mm wide SuperOx tapes showed increased demagnetization rate with the cross-field. The comparison with the calculations revealed that the constant $J_c$ and isotropic $J_c(B)$ dependences are not sufficient for cross-field modelling, and $J_c(B,\theta)$ dependence is necessary. The MEMEP 3D calculations \R{agree with the measurements. Although the accuracy is very high regarding the total trapped field (around 4 \% is the worst of the cases), the demagnetization rate presents deviations as high as 29 \%.} \R{To improve the predictions at higher applied fields, it might be necessary to take the full solid angle dependence, $J_c(B,\theta,\phi)$, into account, which includes macroscopic force-free effects.} The $n(B,\theta)$ dependence showed \R{only} a slight influence on the demagnetization rate.      

\R{The results in this article indicate that the 2D assumption causes a higher error in the demagnetization rate, as defined in equation (3), than the expected deviation from neglecting force-free effects for all ripple field amplitudes. However, for high ripple fields, both features produce a comparable error. Since 2D modeling also inherently neglects force-free effects, both errors accumulate, which might cause an error as large as 60 \% (at the 20th cycle and large ripple fields).}

In conclusion, we have shown that 3D modelling can qualitatively predict cross-field demagnetization in stacks of tapes, contrary to previous 2D modelling. This qualitative study has been possible thanks to the computing efficiency and parallelization of the MEMEP 3D method. The analysis here suggests that force-free effects may be important in the measured samples, pointing out the \R{interest} of $J_c(B,\theta,\phi)$ measurements over the whole solid angle range.


\section*{Acknowledgements}

We acknowledge M vojen\v ciak for the critical current and power-law exponent measurements and A Patel for discussions. The authors acknowledge the financial support by the European Union's Horizon 2020 research innovation program under grant agreement No 7231119 (ASuMED consortium). M.K. and E.P. acknowledge the use of computing resources provided by the project SIVVP, ITMS 26230120002 supported by the Research $\&$ Development Operational Programme funded by the ERDF, the financial support of the Grant Agency of the Ministry of Education of the Slovak Republic and the Slovak Academy of Sciences (VEGA) under contract No. 2/0097/18, and the support by the Slovak Research and Development Agency under the contract No. APVV-14-0438. 




\bibliographystyle{unsrt}

\end{document}